\documentclass[twocolumn,floatfix,prb,showpacs,color,superscriptaddress,epsfig]{revtex4}
\usepackage{graphicx}
\usepackage{amsfonts}
\usepackage{amsmath}
\usepackage{amssymb}
\usepackage{epsfig}
\begin{document}
\title{Nearest neighbor exchange in Co- and Mn-doped ZnO}
\author{T. Chanier}
\affiliation{Laboratoire Mat\'eriaux et Micro\'electronique de
Provence, Facult\'e St.\ J\'er\^ome, Case 142, F-13397 Marseille
Cedex 20, France}
\author{M. Sargolzaei}
\affiliation{Leibniz-Institut f\"ur Festk\"orper- und
Werkstoffforschung Dresden, P.O.Box 270116, D-01171 Dresden,
Germany}
\author{I. Opahle}
\affiliation{Leibniz-Institut f\"ur Festk\"orper- und
Werkstoffforschung Dresden, P.O.Box 270116, D-01171 Dresden,
Germany}
\author{R. Hayn}
\affiliation{Laboratoire Mat\'eriaux et Micro\'electronique de
Provence, Facult\'e St.\ J\'er\^ome, Case 142, F-13397 Marseille
Cedex 20, France}
\author{K. Koepernik}
\affiliation{Leibniz-Institut
f\"ur Festk\"orper- und Werkstoffforschung Dresden, P.O.Box 270116,
D-01171 Dresden, Germany}
\date{\today}

\begin{abstract}
We calculate the magnetic interactions between two nearest neighbor
substitutional magnetic ions (Co or Mn) in ZnO by means of density
functional theory and compare it with the available experimental
data. Using the local spin density approximation we find a
coexistence of ferro- and antiferromagnetic couplings for ZnO:Co, in
contrast to experiment. For ZnO:Mn both couplings are
antiferromagnetic but deviate quantitatively from measurement. That
points to the necessity to account better for the strong electron
correlation at the transition ion site which we have done by
applying the LSDA+$U$ method. We show that we have to distinguish
two different nearest neighbor exchange integrals for the two
systems in question which are all antiferromagnetic with values
between -1.0 and -2.0 meV in reasonable agreement with experiment.
\end{abstract}



\maketitle

\section{Introduction}

The manipulation of the electronic spin for information processing
gives rise to many advantages and would open the way to new
applications. This field of spintronics has need of a
semiconducting, ferromagnetic material at room temperature. In that
respect, several observations of room temperature ferromagnetism in
ZnO:Co \cite{Lee02,Prellier} or ZnO:Mn \cite{Sharma} had been
reported. These experimental findings were partly based on
theoretical predictions using density functional calculations in the
local spin density approximation (LSDA). \cite{Sato,Sharma} However,
they are not at all confirmed. Whereas the ferromagnetic thin films
were mainly produced by laser ablation,\cite{Prellier,Sharma} or by
the sol-gel method,\cite{Lee02}other samples fabricated by precursor
deposition,\cite{Lawes05} molecular beam epitaxy \cite{Sati} or
powder samples \cite{Yoon} showed no signs of ferromagnetism. In
that contradictory situation we propose to study very carefully the
magnetic interaction between two nearest neighbor substitutional
magnetic ions (Co or Mn) in the dilute limit without any codoping
effect. Namely, the numerical values of these exchange couplings are
already quite well known by susceptibility \cite{Yoon} or
magnetization step measurements \cite{Gratens} which allow a careful
check of the LSDA results. Therefore, we present accurate full
potential band structure calculations within density functional
theory. We find a rather remarkable discrepancy between the measured
data and the LSDA results which can be considerably reduced by
taking into account more properly the correlation effects of the
transition metal ions within the LSDA+$U$ method.

First LSDA studies found ZnO:Co to be ferromagnetic, but ZnO:Mn
antiferromagnetic.\cite{Sato} Pseudopotential calculations on large
supercells\cite{Lee04,Sluiter} which were performed later on, could
specify the different couplings more in detail. They found a
competition between ferromagnetic and antiferromagnetic interactions
in ZnO:Co \cite{Lee04,Sluiter} which we confirm and argued for the
necessity of additional electron \cite{Lee04,Sluiter} or hole doping
\cite{Sluiter,Spaldin} to stabilize ferromagnetic order. (Such an
additional doping will not be studied here.) But the presence of
ferromagnetic and antiferromagnetic couplings at the same time would
lead to a very small Curie-Weiss constant in contrast to the
observed one which is clearly antiferromagnetic.
\cite{Yoon,Sati,Lawes05} Even more clear is the contradiction
between the measured data and the LSDA results for ZnO:Mn:
magnetization step measurements \cite{Gratens} lead to
antiferromagnetic exchange integrals of -1.56 and -2.08 meV for the
two nearest neighbor positions possible, although the experiment
does not allow to assign these values to certain bonds in an
unambiguous way. Transforming the energy differences into numerical
values of exchange integrals (which was not done in Refs.\
\onlinecite{Lee04,Sluiter}) we will show that LSDA overestimates the
experimental results considerably. After having stated the
discrepancy between LSDA and experiment, we show that the LSDA+$U$
method can cure these deficiencies and leads to reasonable exchange
couplings. The importance of the LSDA+$U$ approximation has also
recently been pointed out for the related system ZnSe:Mn.
\cite{Sandratskii}

\section{Supercell calculations}

To determine the nearest neighbor exchange couplings we performed
several supercell calculations. ZnO crystallizes in the hexagonal
wurtzite structure (space group P63mc) with the lattice parameters
$a=3.2427$ \AA \ and $c=5.1948$ \AA .\cite{Sabine} We consider here
pure substitutional defects and neglect the influence of lattice
relaxations. Due to the wurtzite structure there are two
crystallographically different nearest neighbor positions: the
in-plane nearest neighbor within the plane perpendicular to the
hexagonal axis ${\bf c}$ and the out-of-plane nearest neighbor.
Their magnetic couplings were studied by using 4 different
supercells (A,B,C, and D). Each supercell is formed by multiples of
the primitive lattice vectors ${\bf a}$, ${\bf b}$, and ${\bf c}$,
like the $2 \times 2 \times 1$ supercell A, shown in Fig.\ 1. The
supercells A and C probe the in-plane nearest neighbor exchange
$J_{in}$ (see Fig.\ \ref{fig1b}) by a chain of Co (or Mn) impurities
(supercell A) or an isolated pair ($3 \times 2 \times 1$ supercell
C). The out-of-plane exchange $J_{out}$ is probed by the $2 \times 2
\times 1$ supercell B (chain) and the $2 \times 2 \times 2$
supercell D (pair). The so defined supercells A-D coincide with
those used in Ref.\ \onlinecite{Lee04}.

\begin{figure}
\includegraphics[scale=0.35,angle=0]{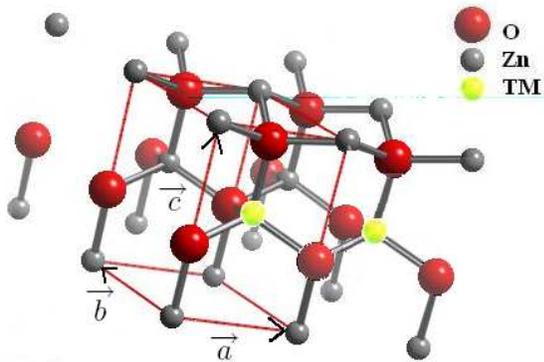}\\
\caption{(Color online) Crystal structure of the supercell A. (TM =
Co, Mn)} \label{fig1}
\end{figure}

\begin{figure}
\includegraphics[scale=0.35,angle=0]{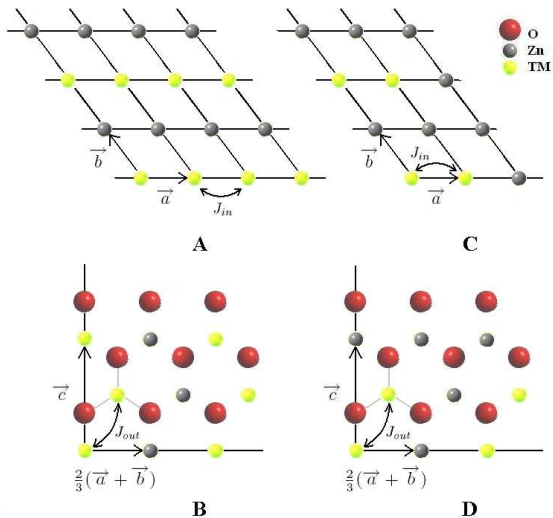}\\
\caption{(Color online) ($\vec{a}$,$\vec{b}$) plane corresponding to
supercells A and C as well as ($\frac{2}{3}(\vec{a}+\vec{b})$,
$\vec{c}$) plane corresponding to supercells B and D, with the
definition of in-plane and out-of-plane exchange constants $J_{in}$
and $J_{out}$, respectively. (TM = Co, Mn)}  \label{fig1b}
\end{figure}

The supercell calculations were performed using the full-potential
local-orbital (FPLO) band structure scheme. \cite{Koepernik} In the
FPLO method a minimum basis approach with optimized local orbitals
is employed, which allows for accurate and efficient total energy
calculations. For the present calculations we used the following
basis set: Zn, Co(Mn) $3s3p$:$4s4p3d$, O $2s2p$;$3d$. The inclusion
of the transition metal $3s$ and $3p$ semicore states into the
valence was necessary to account for non-negligible core-core
overlap, and the O $3d$ states were used to improve the completeness
of the basis set. The site-centered potentials and densities were
expanded in spherical harmonic contributions up to $l_{max}=12$.

The exchange and correlation potential was treated in two different
ways. First, the local spin-density approximation (LSDA) was used in
the parametrization of Perdew and Wang. \cite{Perdew} However, as
will be shown more in detail below, this approximation has severe
deficiencies in the present case. The Co(Mn) $3d$ states are in
reality more localized than in the LSDA calculation. This
correlation effect was taken into account by using the FPLO
implementation of the LSDA+$U$ method in the atomic limit scheme.
\cite{Anisimov91,Eschrig} The convergence of the total energies with
respect to the ${\bf k}$-space integrations were checked for each of
the supercells independently. We found that $8 \times 8 \times 8
=512$ ${\bf k}$-points were sufficient in all cases, and this
parameter was used in the calculations reported below.

\section{Exchange couplings}

The Zn$^{2+}$ ion in ZnO has a completely filled $3d$ shell and,
correspondingly, no magnetic moment. If Zn is replaced by Co or Mn,
the valence 2+ is not changed, which is also proved by our
bandstructure results below. It means that these substitutional
impurities provide no charge carriers. The configuration of
Co$^{2+}$ is $d^7$ and that of Mn$^{2+}$ is $d^5$. Therefore, they
have a spin $S=3/2$ or $S=5/2$, correspondingly. The tetrahedral and
trigonal crystal fields, together with the spin-orbit coupling, lead
to a magnetic anisotropy. But they are not large enough to
destabilize the high-spin states in the given cases, as supported by
electron paramagnetic resonance and magnetization measurements (see
Refs.\ \onlinecite{Estle61,Jedrecy04,Sati} for ZnO:Co and Ref.\
\onlinecite{Schneider62} for ZnO:Mn).

The present work is devoted to determine the dominant exchange
couplings between two localized magnetic ions. It can be expected
that the dominant couplings occur between nearest neighbor
impurities, each carrying a local spin ${\bf S}_i$. Then, the
Heisenberg  Hamiltonian for a localized pair of spins is given by
\begin{equation}
H=-2J{\bf S}_i {\bf S}_j \; .
\end{equation}
The corresponding total energies for ferromagnetic (FM) and
antiferromagnetic (AFM) arrangements of the two spins are
\begin{eqnarray}
E_{FM} &=& -J\left[ S_T(S_T+1) - 2S(S+1) \right] \; , \nonumber \\
E_{AFM} &=& J \left[ 2S(S+1) \right] \; ,
\end{eqnarray}
with the total spin $S_T=2S$ of two parallel spins $S$. This leads
to the energy difference between the FM and AFM states per magnetic
ion:
\begin{equation}
\Delta E = \frac{E_{FM}-E_{AFM}}{2} = -\frac{J}{2}S_T(S_T+1) \; ,
\label{pair}
\end{equation}
with $S_T=3$ or 5 for Co or Mn. That energy difference can be
compared with the corresponding energy differences of isolated pairs
in the supercells C and D. The supercells A and B, however,
correspond to chains of magnetic ions. For our purpose, it is
sufficient to use an approximative expression for the energy
difference between FM and AFM states of a Heisenberg chain. Then,
each magnetic ion has two nearest neighbor magnetic ions which
doubles the previous energy difference (\ref{pair}):
\begin{equation}
\Delta E = \frac{E_{FM}-E_{AFM}}{2} = - J S_T(S_T+1) \; .
\end{equation}
That result can also be calculated more explicitly by comparing the
energies of the FM and AFM states of a Heisenberg chain decomposed
into a series of pairs.

It is remarkable that the so defined exchange couplings are
experimentally measurable. They will be denoted here as $J_{in}$ and
$J_{out}$ for the in-plane and out-of-plane nearest neighbors,
respectively. The most precise measurements had been performed using
magnetization steps \cite{Gratens} for the case of ZnO:Mn which
leads to different values for $J_{in}$ and $J_{out}$. We are not
aware of such measurements for ZnO:Co. An average value of $J_{in}$
and $J_{out}$ is however accessible by susceptibility measurements.
\cite{Yoon}

\section{Results}

\subsection{ZnO:Co}

\begin{figure}
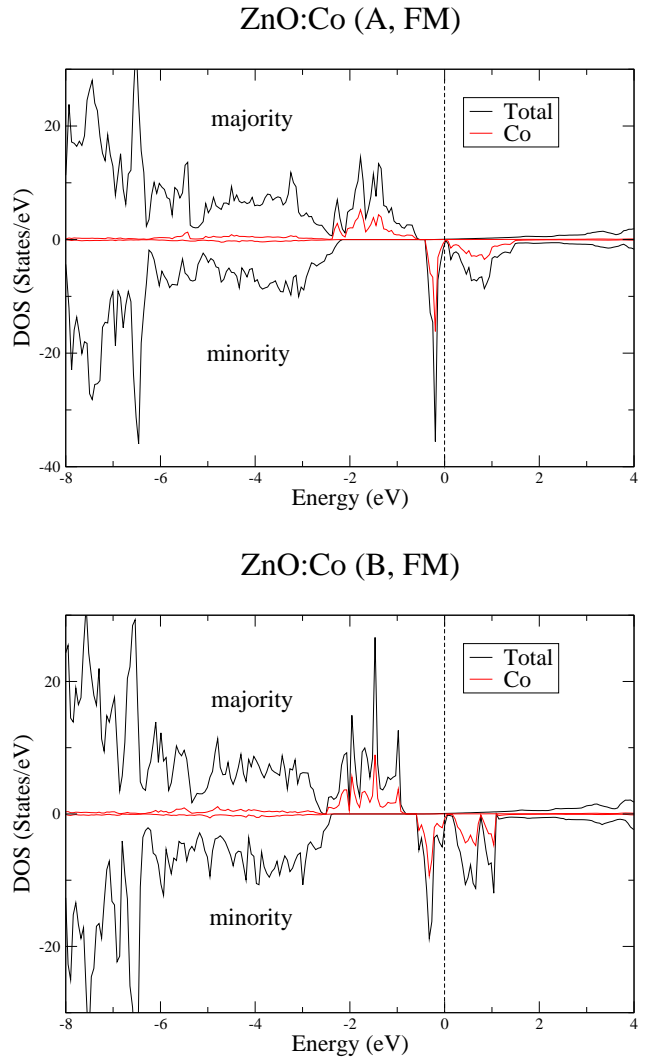

\includegraphics[scale=0.35,angle=0]{fig3.eps}\\[0.5cm]
\includegraphics[scale=0.35,angle=0]{fig4.eps}\\
\caption{(Color online) Comparison of total and partial DOS (for 1
Co atom) for supercells A and B, ZnO:Co, in the FM case obtained by
LSDA.} \label{fig2}
\end{figure}

The density of states (DOS) of the FM solution (for supercells A and
B) is shown in Fig.\ \ref{fig2}. Its main features agree with
previous calculations. \cite{Lee04} The minority and majority Co
$3d$ states are found to lie mainly in the gap between the valence
band of predominantly oxygen $2p$ character and the Zn $4s$-$4p$
conduction band. The Zn $3d$ states are located at about -7 eV at
the bottom of the valence band, deep below the Fermi level. The
occupation of the Co $3d$ level is close to $3d^7$ and the
magnetization $M_s^{FM}(Co)=2.6 \mu_B$, i.e.\ rather close to
$S=3/2$. The majority Co $3d$ states are located just above the
oxygen valence band. There is a small hybridization of the minority
$3d$ level with the conduction band which makes the material
half-metallic in the LSDA (see also
Refs.~\onlinecite{Lee04,Sluiter}). However, this half-metallic
character would correspond to a partial electron doping and is an
artifact of the LSDA solution.

Already the gap of pure ZnO (experimental gap: 3.3 eV) is
underestimated by LSDA (FPLO leads to a gap of 1.4 eV). To study the
3$d$ levels of an isolated Co impurity we calculated the electronic
structure of a CoZn$_7$O$_8$ supercell by LSDA. The corresponding
DOS of the FM solution is rather close to that one shown in Fig.\
\ref{fig2}, but the band structure (not shown) allows a better
analysis of the impurity levels, free of hybridization effects
between neighboring impurities. For instance, the crystal field
splitting of the Co $3d$ levels is clearly visible at the
$\Gamma$-point. In each of the spin channels the two-fold
degenerated $e_g$ levels are situated below the $t_{2g}$ orbitals,
for majority  spin by about 0.5 eV. The $e_g$ levels remain
degenerate in the trigonal case, whereas the $t_{2g}$ ones split
into a lower singlet and an upper doublet. This trigonal splitting
is of the order of 0.2 eV. Due to hybridization effects, the Co $3d$
impurity band is considerably broader in Fig.\ \ref{fig2} than for
CoZn$_7$O$_8$ by roughly a factor of three for majority spin. That
shows that hybridization and crystal field splitting are of the same
order of magnitude for two neighboring Co impurities in ZnO.

The differences of total energies between FM and AFM solutions give
the corresponding exchange couplings in the way described above. The
values for in-plane and out-of-plane exchange $J_{in}$ and $J_{out}$
are collected in Tables I and II. Like in the previous
pseudopotential calculations \cite{Lee04,Sluiter} the in-plane
exchange is antiferromagnetic, but the out-of-plane exchange
ferromagnetic. The differences between supercells A and C (or
between B and D, correspondingly) arise due to finite size effects
or deviations from the Heisenberg model. Qualitatively, all
available LSDA energy differences agree among each other and with
our FPLO results (see Table III). But there are rather remarkable
numerical deviations between the different methods. Please note,
that the previous authors \cite{Lee04,Sluiter} did not convert the
LSDA energy differences into exchange couplings. That we have done
to allow the comparison with experimental data.

\begin{table}
\begin{tabular}{c|c|c|c|c|c|c|c}
Cell & U[eV] & $\Delta$ E[meV/Co] & $J_{in}$[meV] &  $M_s^{AF}(\rm
Co)$ & $M_s^{FM}(\rm Co)$\\\hline
A & 0 & 22 & -1.8 & 2.49 & 2.60 \\
C & 0 & 16 & -2.6 & 2.54 & 2.60 \\
A & 6 & 24 & -2.0 & 2.81 & 2.82 \\
A & 8 & 36 & -1.5 & 2.86 & 2.86\\
\hline
\end{tabular}
\caption{Calculated in-plane exchange $J_{in}$ for ZnO:Co using LSDA
($U=0$) and LSDA+$U$ ($F^{0}=U\neq0$, $F^{2}=7.9$~eV and
$F^{4}=5.0$~eV) for the supercells A and C. Also given are the
corresponding energy differences per Co ion and the magnetic
moments.}
\end{table}

\begin{table}
\begin{tabular}{c|c|c|c|c|c|c|c}
Cell & U[eV] & $\Delta$ E[meV/Co] & $J_{out}$[meV] & $M_s^{AF}(\rm
Co)$ & $M_s^{FM}(\rm Co)$\\\hline
B & 0 & -31 & 2.6 & 2.52 & 2.60\\
D & 0 & -14 & 2.4 & 2.56 & 2.60\\
B & 6 & 12 & -1.0 & 2.81 & 2.82\\
B & 8 & 12 & -1.0 & 2.86 & 2.87\\
\hline
\end{tabular}
\caption{Calculated out-of-plane exchange $J_{out}$ for ZnO:Co using
LSDA ($U=0$) and LSDA+$U$ ($F^{0}=U\neq0$, $F^{2}=7.9$~eV and
$F^{4}=5.0$~eV) for the supercells B and D. Also given are the
corresponding energy differences per Co ion and the magnetic
moments.}
\end{table}

The competition between FM and AFM nearest neighbor exchange in
ZnO:Co is in contrast to experimental results
\cite{Yoon,Sati,Lawes05} which show dominantly AFM couplings. Other
problems of the LSDA solution are the following: (i) the
semi-metallic character, (ii) the insufficient localization of the
Co $3d$ states, and (iii) the position of the impurity 3$d$ levels.
Namely, photoemission spectroscopy shows them as deep impurity
levels close to the valence band.\cite{CondmatWi} But the
experimental energy difference to the top of the valence band of
only 0.4 eV is smaller than the corresponding distance of the center
of gravity of the Co $3d$ level (about 1 eV for majority spin).

\begin{figure}
\includegraphics[scale=0.35,angle=0]{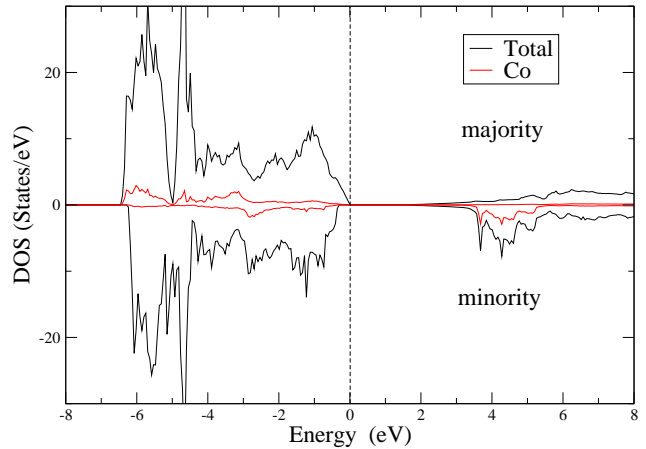}\hskip1cm
\caption{(Color online) DOS of ZnO:Co, supercell A, in the FM case,
calculated with LSDA+$U$.}\label{fig3}
\end{figure}

So, we should look for a theoretical method which takes into account
correlation effects more properly. One has to distinguish the
correlation in valence and conduction band leading to the incorrect
gap value and the correlation effects in the Co $3d$ orbitals. The
first effect might be repaired by the GW approximation,\cite{GW} or
including the self interaction correction (SIC) as proposed in Ref.
\onlinecite{Filipeti}. But we do not expect that it would
considerably improve the exchange couplings. The correlation in the
Co $3d$ shell will be taken into account in our present work by the
LSDA+$U$ scheme using the atomic limit functional. \cite{Czyzyk} We
also tried the "around mean field" version of LSDA+$U$ which gives
similar results as reported below for ZnO:Co, but which leads to no
improvements for ZnO:Mn due to the known peculiarities of this
functional for the $d^5$ configuration. \cite{Czyzyk,Eschrig} The
parameters for ZnO:Co have been chosen similar to those for CoO,
\cite{Anisimov91} namely the Slater parameters $F^2$ and $F^4$ such
that the Hund's rule exchange $J^H=(F^2+F^4)/14=0.92$ eV and the
ratio $F^4/F^2=0.625$ are close to ionic values \cite{Anisimov93}
which leads to $F^2=7.9$ eV and $F^4=5.0$ eV. The parameter $F^0=U$
is less well known since it is more affected by screening effects.
We found that a value of at least 3 eV is necessary to stabilize an
insulating solution. In the region of realistic $F^0$ parameters its
influence on the values $J_{in}$ and $J_{out}$ is small as indicated
by the comparison of results for $F^0=6$ and 8 eV in Tables I and
II.

The DOS (Fig.\ \ref{fig3}) shows then clearly an insulating state
and the occupied Co levels are much closer to the valence band than
in LSDA in better agreement with photoemission data.
\cite{CondmatWi} The gap value ($E_{g}\approx 0.7$ eV for the FM
case, $F^0=6$ eV) is not much improved, but that can also not be
expected since we did not change the potentials for oxygen or zinc
$s$-$p$ states. However, it is remarkable that now both exchange
couplings are antiferromagnetic.

We find the in-plane exchange $J_{in}=-2.0$ meV to be larger than
the out-of-plane exchange $J_{out}=-1.0$ meV (for $U=6$ eV).
Experimentally, the antiferromagnetic nearest neighbor exchange was
determined to be $J=-33$ K or -2.8 meV from the high-temperature
Curie-Weiss constant of the magnetic susceptibility,\cite{Yoon}
which exceeds slightly our values.

\begin{table}\label{Tabjcalc}
\begin{tabular}{c|c|c|c|c|c|c|c}
& $J_{in}^{Co}$ & $J_{out}^{Co}$ & $J_{in}^{Mn}$ & $J_{out}^{Mn}$
\\\hline
Ref.\ \onlinecite{Lee04} cells A, B & -2.8 & 0.1 & & \\
Ref.\ \onlinecite{Lee04} cells C, D & -3.5 & 0.2 & & \\
Ref.\ \onlinecite{Sluiter}    & -3.3 & 2.5 & -4.0 & -3.3 \\
\hline
\end{tabular}
\caption{Available theoretical LSDA results for $J_{in}$ and
$J_{out}$ (all data in meV) of ZnO:Co and ZnO:Mn taken from the
literature. The published energy differences were converted into
exchange constants.}
\end{table}

\subsection{ZnO:Mn}
In Fig.\ 5, we show the DOS of ZnO:Mn, calculated with supercell A
and a ferromagnetic arrangement of the Mn moments. As in the case of
ZnO:Co, LSDA yields a metallic solution. The Mn $3d$ shell is
approximately half filled and the exchange splitting between the
centers of gravity of the occupied majority and unoccupied minority
subbands is about 3.5~eV. The total spin moment $4.96\mu_B$/Mn is
close to the expected $S=5/2$ value. The Mn atoms carry a spin
moment of about $4.6\mu_B$, but also their four nearest neighbor O
atoms have a weak induced spin moment. The Mn $3d$ impurity states
are mainly located in the upper part of the ZnO gap and weakly
hybridize with the Zn 4$s$-4$p$ conduction band, which gives the
solution a metallic character. Measurements of the band gap of
ZnO:Mn films~\cite{Cheng03} (see Ref.~\onlinecite{Oezguer05} for an
overview) find a slight blue shift of the absorption edge with a
significant amount of mid gap absorption above 2.5 eV. This is
consistent with a position of the impurity levels around the upper
edge of the valence band in contrast to the LSDA result.

As shown in the lower graph of Fig.\ 5, LSDA+$U$ shifts the highest
occupied Mn $3d$ levels to the top of valence band, so that the
solution becomes insulating. The parameters used in the calculation
are $U=F^0=6$~eV, $F^2=7.4$~eV and $F^4=4.6$~eV, corresponding to
$J^H=0.86$~eV, the value chosen for MnO in
Refs.~\onlinecite{Anisimov91,Anisimov93}. As in the case of ZnO:Co,
the value of the band gap $E_g\approx 0.4$ eV is smaller than the
experimental one (see discussion above), but the position of the Mn
$3d$ impurity levels is considerably improved. Compared to the LSDA
calculation, the partial Mn $3d$ DOS is slightly broadened and the
unoccupied Mn $3d$ minority spin states are shifted further away
from the Fermi level. The total spin moment is now $5\mu_B$
corresponding to an ideal $S=5/2$ situation with the magnetic
contributions almost entirely due to Mn (see Tables IV and V).

In contrast to the case of ZnO:Co, LSDA yields an AFM exchange
coupling for both types of nearest neighbor pairs in ZnO:Mn. This is
in qualitative agreement with the magnetization step measurements of
Ref.~\onlinecite{Gratens}, where values $J_{in}=-2.08$ meV and
$J_{out}=-1.56$ meV have been obtained. However, the LSDA values
$J_{in}=-4.9$ meV and $J_{out}=-4.1$ meV are 2-3 times larger than
the experimental ones. Similar results have also been obtained by
Sluiter {\em et al.}\cite{Sluiter}, who find both couplings to be
strongly AFM (see Tab. III). However, the rather poor quantitative
agreement between the calculated and measured $J$ values indicates
that the Mn impurity levels are not well described within LSDA.
Again, the LSDA+$U$ functional strongly improves the agreement of
the calculations with experimental data. Taking $U=6$~eV, we obtain
$J_{in}=-2.0$ meV and $J_{out}=-1.3$ meV, close to the measured
values. As can be seen in Tables IV and V, the calculated exchange
couplings depend only moderately on the choice of the $U$-parameter.
The larger coupling is obtained for the in-plane pairs despite the
larger distance, as was already assumed in
Ref.~\onlinecite{Gratens}. \

\begin{figure}
\includegraphics[scale=0.35,angle=0]{fig6.eps}\\
\includegraphics[scale=0.35,angle=270]{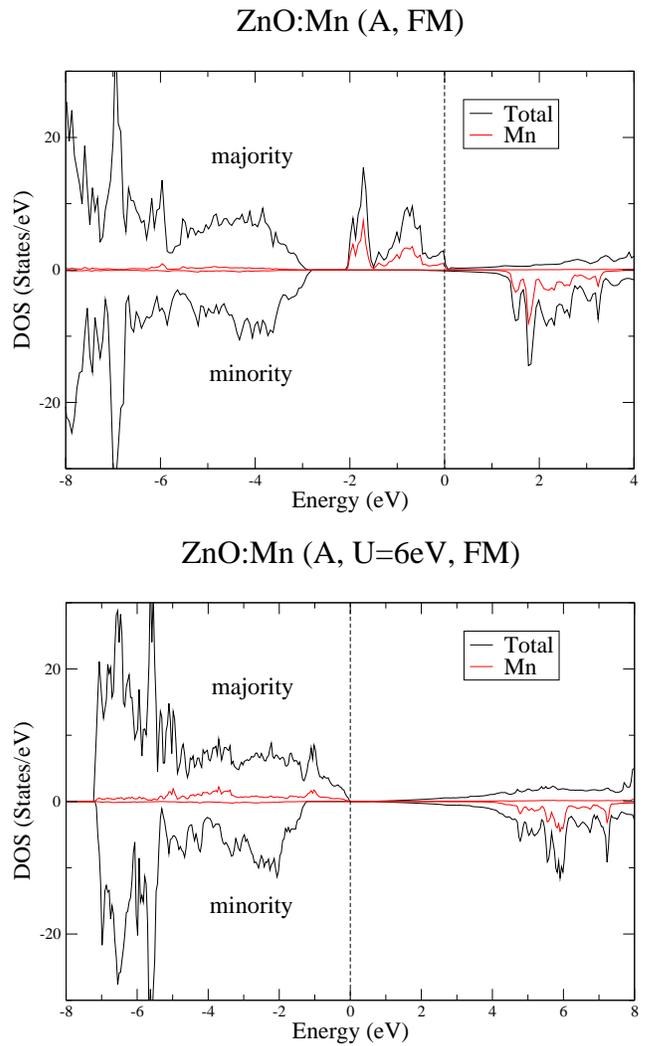}\\
\caption{(Color online) DOS of ZnO:Mn, comparison of LSDA and
LSDA+$U$, supercell A.}\label{fig4}
\end{figure}

\begin{table} \label{Tab:mnzno1}
\begin{tabular}{c|c|c|c|c|c|c|c}
Cell & U[eV] & $\Delta$ E[meV/Mn] & $J_{in}$[meV] &  $M_s^{AF}(\rm
Mn)$ & $M_s^{FM}(\rm Mn)$\\\hline
A & 0 & 147  & -4.9 & 4.52 & 4.62\\
C & 0 & 74 & -4.9 & 4.57 & 4.62\\
A & 6 & 59 & -2.0 & 4.87 & 4.89\\
A & 8 & 48 & -1.6 & 4.94 & 4.94\\
\hline
\end{tabular}
\caption{Calculated in-plane exchange $J_{in}$ for ZnO:Mn using LSDA
($U=0$) and LSDA+$U$ ($F^{0}=U\neq 0$, $F^{2}=7.4$~eV and
$F^{4}=4.6$~eV) for the supercells A and C. Also given are the
corresponding energy differences per Mn ion and the magnetic
moments.}
\end{table}

\begin{table} \label{Tab:mnzno2}
\begin{tabular}{c|c|c|c|c|c|c|c}
Cell & U[eV] & $\Delta$ E[meV/Mn] & $J_{out}$[meV] &  $M_s^{AF}(\rm
Mn)$ & $M_s^{FM}(\rm Mn)$\\\hline
B & 0 & 122  & -4.1 & 4.55 & 4.64\\
D & 0 & 57 & -3.8 & 4.57 & 4.61\\
B & 6 & 40 & -1.3 & 4.88 & 4.89\\
B & 8 & 30 & -1.0 & 4.94 & 4.94\\
\hline
\end{tabular}
\caption{Calculated out-of plane exchange $J_{out}$ for ZnO:Mn using
LSDA ($U=0$) and LSDA+$U$ ($F^{0}=U\neq0$, $F^{2}=7.4$~eV and
$F^{4}=4.6$~eV) for the supercells B and D. Also given are the
corresponding energy differences per Mn ion and the magnetic
moments.}
\end{table}

\section{Discussion}

\begin{figure}
\includegraphics[scale=0.35,angle=0]{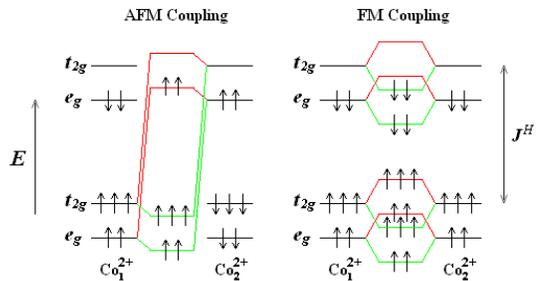}\hskip1cm
\caption{(Color online) Schematic view of hybridization and CF
effects on two close Co $3d$ shells in the LSDA case, comparison
between AFM and FM couplings. The energy levels of majority spin are
lower in energy than minority spin ones due to the Hund's coupling
$J^{H}$.}\label{fig5}
\end{figure}

The reason for the competition of FM and AFM exchange couplings
within LSDA for ZnO:Co is schematically shown in Fig.\ \ref{fig5}.
Let us first consider an isolated Co ion in ZnO. The $3d$ levels are
split by the crystal field (CF) into lower $e_g$ and upper $t_{2g}$
levels. They are also influenced by the exchange splitting between
spin up and spin down electrons caused by the local Hund's rule
exchange. These local energy levels are filled with 7 electrons in
the case of Co. The LSDA-DOS (Figs.\ 3,5) shows that the CF
splitting is smaller than the exchange splitting. Fig.\ \ref{fig5}
presents the hybridization effect on the Co $3d$ energy levels if
the two Co ions come close together. The hybridization leads to the
formation of a pair of bonding and antibonding hybrid orbitals for
each $3d$ energy level $E_i$ of Co. The bonding and antibonding
orbitals have energies $E_i-\Delta E_i$ and $E_i+\Delta E_i$,
correspondingly. Therefore, the complete filling of these two
orbitals does not lead to an energy gain, but a partial filling
does. In such a way, the energy gain for an AFM arrangement of spins
is evident. For a FM arrangement, the energy gain is only possible
by a crossing of the $e_g$ and $t_{2g}$ levels for minority spin.
This competition between the FM and the AFM energy gain is
apparently not identical for in-plane and out-of-plane exchange,
leading to different signs of the exchange couplings. However, as
already discussed, that is an artifact of the LSDA solution.

\begin{figure}
\includegraphics[scale=0.35,angle=0]{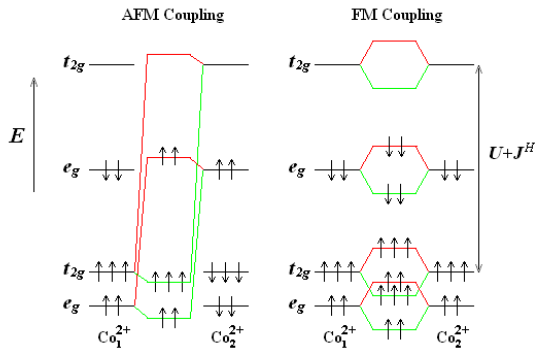}\hskip1cm
\caption{(Color online) Schematic view of hybridization and CF
effects on two close Co $3d$ shells in the LSDA+$U$ case, comparison
between AFM and FM couplings. The energy levels of majority spin are
lower in energy than minority spin ones due to the Hund's coupling
$J^{H}$. The occupied energy levels are lower in energy than
unoccupied ones due to the $3d$ shell correlation effect
$U$.}\label{fig6}
\end{figure}

In LSDA+$U$ (Fig.\ \ref{fig6}) the unoccupied minority spin energy
levels are much higher than the occupied ones. Therefore, the
crossing of minority $e_g$ and $t_{2g}$ energy levels, and also the
FM energy gain, is not possible. As a consequence, one finds an AFM
superexchange coupling in ZnO:Co independent of the geometrical
configuration.

The situation for ZnO:Mn is different. In that case, only the
majority spin is completely filled with 5 electrons and the minority
spin is nearly empty in LSDA, and completely empty in LSDA+$U$.
Since the exchange splitting is larger than the CF splitting, there
is no energy gain possible for a FM arrangement of impurity spins
neither in LSDA nor in LSDA+$U$. The distance between occupied and
unoccupied energy levels increases in LSDA+$U$. Therefore, the
energy gain is reduced and leads to exchange couplings which are
much closer to the experimental values obtained by magnetization
step measurements than those obtained by LSDA. It should be noted
that our calculation confirms also the assignment of Ref.\
\onlinecite{Gratens} that $J_{in}$ corresponds to the largest
coupling.

Of course, there are still numerical error sources on the exchange
couplings $J_{in}$ and $J_{out}$ which were calculated by LSDA+$U$.
First of all, we should note the poor knowledge of correlation
parameters $U$, $F^2$, and $F^4$, which influences the results.
Second, there might still be finite size effects due to the specific
form of the supercells chosen. And finally, also a small basis set
dependence of the FPLO method cannot be excluded. All together, we
would estimate an upper error of about $\pm 30$ per cent for the
calculated exchange couplings.

\section{Conclusions}

In the dilute limit, nearest neighbor pairs of Co and Mn-impurities
in ZnO have antiferromagnetic exchange couplings. That is the result
of theoretical calculations which take into account the electron
correlations in the impurity $3d$ shell properly, and is in
agreement with the experimental results. This AFM nearest neighbor
exchange excludes ferromagnetism for pure substitutional Co or Mn
defects in ZnO in the dilute limit. The observed FM in ZnO:Co and
ZnO:Mn should have a different origin. There are several proposals
in the literature like secondary phases \cite{Garcia05} or cation
vacancies or other defects.\cite{Venkarsan05,Park05}

The LSDA predictions might be misleading and should be considered
with care since they do not correctly take into account the
localized character of the transition metal impurities. On the
contrary, the LSDA+$U$ values are in good agreement with
experimental exchange constants derived from magnetization step
measurements and high-temperature susceptibility
data.\cite{Gratens,Yoon} So our study puts considerable doubts on
the value of pure LSDA predictions (as published for instance in
Refs. \onlinecite{Sharma},\onlinecite{Sluiter}), at least in the
case without additional electron or hole doping.


\section{acknowledgements}
We thank Laurent Raymond and Ulrike Nitzsche for help with the
numerical calculations and the NATO science division grant (grant
CLG 98 1255) for financial support. We thank also Anatoli Stepanov,
Pascal Sati and Roman Kuzian for useful discussions.

\end{document}